\documentclass[conference]{IEEEtran}
\IEEEoverridecommandlockouts

\usepackage[utf8]{inputenc}

\usepackage{cite}
\usepackage{caption}
\usepackage{subcaption}
\usepackage[implicit=false,bookmarks=false]{hyperref}
\usepackage{amsmath,amssymb,amsfonts}
\usepackage{algorithmic}
\usepackage{graphicx}
\usepackage{textcomp}
\usepackage{xcolor}
\usepackage{mathtools}
\usepackage{booktabs}
\usepackage{siunitx}
\usepackage{multirow}
\usepackage[none]{hyphenat}
\usepackage{lipsum}
\usepackage{multicol}
\usepackage{stfloats}
\usepackage{url}
\usepackage{balance}

\title{Enabling a Cooperative Driver Messenger System for Lane Change Assistance Application}
\author{Ghayoor Shah$^*$, Shahriar Shahram$^*$, Yaser Fallah$^*$, Danyang Tian$^\dagger$, Ehsan Moradi-Pari$^\dagger$\\
$^*$Connected and Autonomous Vehicle Research Lab (CAVREL),\\
University of Central Florida, Orlando, FL\\
$^\dagger$Honda Research Institute, Ann Arbor, MI\\
\{gshah8, shahram\}@knights.ucf.edu, yaser.fallah@ucf.edu, \{danyang\_tian, emoradipari\}@honda-ri.com
}
\date{August 2021}

\begin{document}

\maketitle

\begin{abstract}
Sensor data and Vehicle-to-Everything (V2X) communication can greatly assist Connected and Autonomous Vehicles (CAVs) in situational awareness and provide a safer driving experience. While sensor data recorded from devices such as radar and camera can assist in local awareness in the close vicinity of the Host Vehicle (HV), the information obtained is useful solely for the HV itself. On the other hand, V2X communication can allow CAVs to communicate with each other and transceive basic and/or advanced safety information, allowing each CAV to create a sophisticated local object map for situational awareness. This paper introduces a point-to-point Driver Messenger System (DMS) that regularly maintains a local object map of the HV and uses it to convey HV's Over-the-Air (OTA) Driver Intent Messages (DIMs) to nearby identified Target Vehicle(s) (TV(s)) based on a list of pre-defined common traffic applications. The focus of this paper is on the lane change application where DMS can use the local object map to automatically identify closest TV in adjacent lane in the direction of HV's intended lane change and inform
the TV via a DIM.
Within DMS, the paper proposes a TV recognition algorithm for lane change application that utilizes the HV's Path History (PH) to accurately determine the closest TV that could potentially benefit from receiving a DIM from HV.
Finally, DMS is also shown to act as an advanced warning system by providing extra time and space headway measurements between the HV and TVs upon a number of simulated lane change scenarios.
\end{abstract}


\section{Introduction}
Connected and Autonomous Vehicles (CAVs) can utilize sensor data and Vehicle-to-Everything (V2X) communication to create situational awareness and 
make driving decisions.
While CAVs can rely on data from sensors such as camera and radar for local situational awareness in the vicinity of the Host Vehicle (HV), there can be driving situations desiring a larger local object map encompassing extremely close vehicles along with vehicles that may be comparatively distant but still important for scenario-specific situational awareness. Secondly, there can arise numerous driving conditions where even an elementary level of communication and intention sharing among vehicles could assist in a safer driving experience, something which is not available at all with solely using sensor data.

To resolve this, V2X communication can be used where CAVs can communicate their safety information in the form of Basic Safety Messages (BSMs). This form of low-latency communication can allow vehicles to obtain safety information of both close- and far-range vehicles without relying on the perfect functionality of sensors at all times. In this regard, two main communication technologies, namely Dedicated Short-Range Communication (DSRC) \cite{jkenney:dsrcmain} and Cellular Vehicle-to-Everything (C-V2X) \cite{3gpp:3gppmain}, have been extensively studied for a plethora of different driving scenarios and traffic densities.

Based on the NHTSA report conducted in 2015 \cite{singh2015critical}, around 95\% of the estimated two million vehicle crashes in the U.S from 2005 to 2007 were a consequence of human error, out of which around 35\% were due to driver decision error. Given this report and others \cite{national2012traffic}, it is clear that V2X communication 
can be used to assist drivers in improving driving decisions and reducing the number of hazardous driving scenarios by providing better situational awareness. In this regard, numerous studies have focused on the design \cite{ahmed2011vehicle}, scalability \cite{shah2019real}, \cite{shah2020rve}, and efficacy of V2X communication for Cooperative Vehicle Safety (CVS) applications \cite{tian2016evaluating}, \cite{moradi2013modeling}, \cite{moradi2014hybrid}.
\cite{moradi2015co}.  


\begin{figure}[t]
\centerline{\includegraphics[trim=10 35 25 0,clip,width=.495\textwidth]{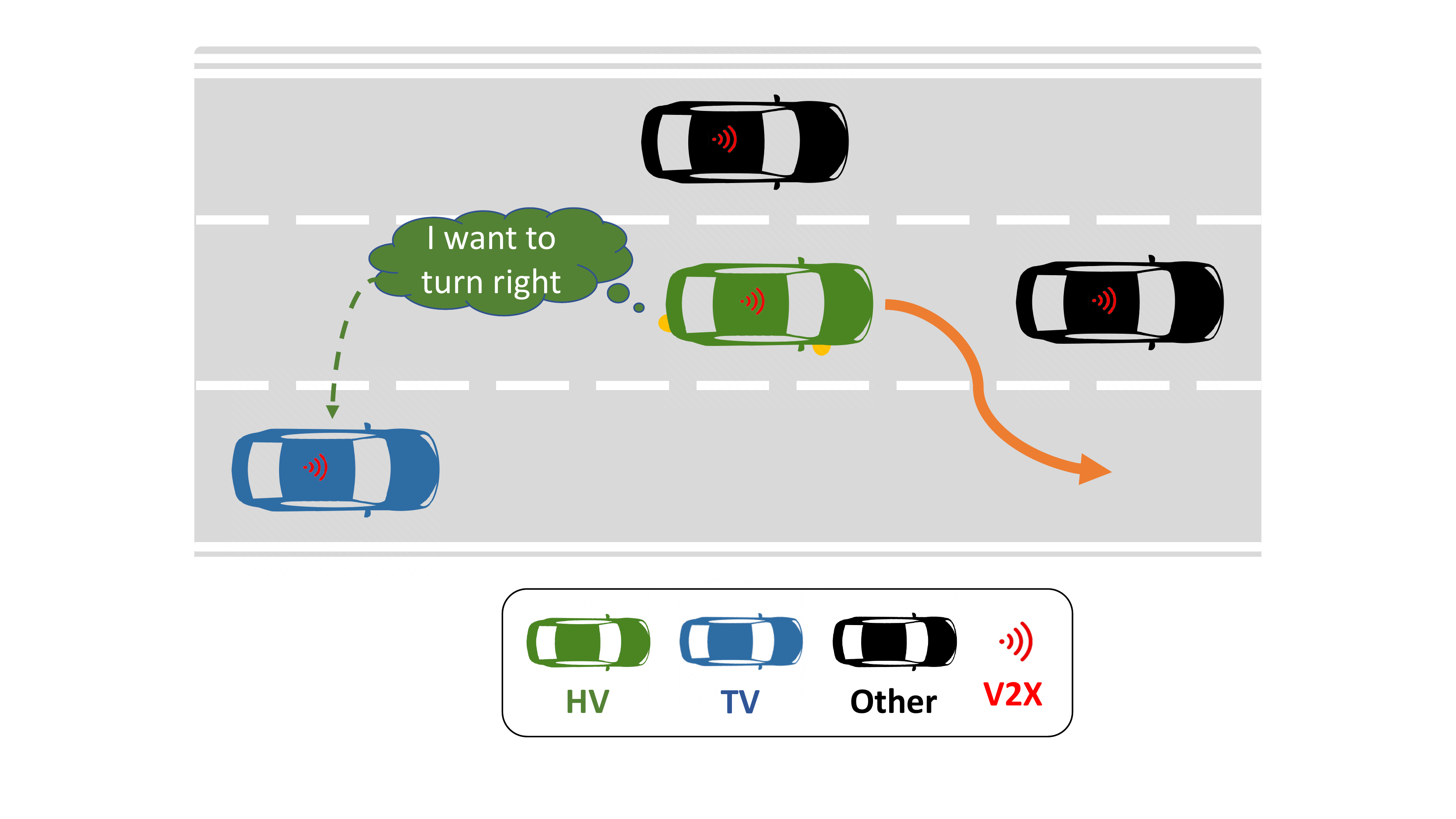}}
\caption{HV conveying lane change intention to TV using DMS}
\label{figlc_scenario}
\end{figure}

In this paper, a point-to-point Driver Messenger System (DMS) is designed and implemented to facilitate CAVs in avoiding risky driving situations.
The goal is to focus on common traffic applications where any conflict or ambiguity among drivers can be resolved/negotiated using V2X communication before a situation can grow into a potentially near-collision scenario. 
DMS uses sensor data and BSMs received from nearby CAVs to update the local object map of the HV. The updated map is then used by DMS to automatically identify one or more of the pre-defined traffic applications. After application detection, DMS identifies potential Target Vehicle(s) (TV(s)) that can benefit from the application-specific information pertaining to the HV. As the final step, DMS sends the corresponding Over-the-Air (OTA) Driver Intent Message (DIM) to the TV(s) that can be viewed by the TV driver through a Driver Vehicle Interface (DVI).
As opposed to CVS systems \cite{ahmed2011vehicle}, \cite{thorn2018framework} where any safety application information is limited to the HV itself, DMS within HV is designed to convey DIMs to TVs, thereby achieving cooperative situational awareness.  

This paper considers the lane change application within DMS
and proposes a TV recognition algorithm that uses HV Path History (PH) in accordance with \cite{sae2016board} to accurately identify TVs that can receive DIM from HV during a lane change scenario.
Figure \ref{figlc_scenario} shows a typical lane change scenario where the HV utilizes DMS to detect and inform the TV regarding its intent to turn right. 
This study shows the effective usage of the lane change application within DMS for both straight and curved roads.
It should be noted that the lane change application within DMS considered here is different from the Lane Change Warning (LCW)/ Blind Spot Warning (BSW) cooperative safety applications within CVS \cite{ahmed2011vehicle}, \cite{thorn2018framework}. LCW/BSW applications consider scenarios where trailing vehicles are either in the blind-spot region of the HV or about to enter the blind-spot region as per the speed difference between the trailing vehicles and HV. On the other hand, lane change application within DMS considers trailing vehicles beyond the blind-spot region and regardless of the speed difference with the HV.

The rest of the paper is organized as follows. Section II discusses the background pertaining to this study along with the related works in literature. Section III highlights the DMS overview with an explanation concerning all the major components including the proposed TV recognition algorithm. In Section IV, the analysis and results of DMS usage in lane change are presented using a real-world dataset and a traffic simulator, and finally, sections V and VI provide conclusions and future works related to this study, respectively.

\section{Background}
Existing studies and government survey data show that aggressive driving behaviors are prevalent among U.S. drivers \cite{aaaroadrage}, including aggressively switching lanes or cutting in front of another vehicle \cite{aaaprevalence}. 
As per police report data regarding fatal crashes in Fatality Analysis Reporting System (FARS), around 1.3\% of the reported fatal crashes in 2019 were attributed to improper or erratic lane changing \cite{fars2019}.
The inability to communicate on road is one of the major factors contributing to selfish driving, road hostility and even road violence \cite{smithsafety}. In this regard, point-to-point messages based on vehicle connectivity can be used to connect drivers and avoid potential on-road misunderstandings and conflicts, thereby potentially improving on-road safety. There are some existing studies on two-way driver-to-driver communication in the process of overtaking to control the vehicles' relative speed \cite{sohr2009radio}, and transmitting DIMs in a cooperating driving scenario \cite{rech2021method}. One relevant study was proposed in 2017 where researchers conducted the driver-to-driver communication concept evaluation in batch via driving simulator for different scenarios, and showed overall acceptance and potential safety benefits of the driver-to-driver communication technology concept \cite{rajab2017driver}. 
However, to the authors' best knowledge, no known literature exists on the particular DMS introduced in this paper and also on the proposed TV recognition algorithm for lane change application.

There are some related studies on TV Classification (TC) \cite{sae2016board}, \cite{lee2005evaluation}, \cite{kim2016target}, however they cannot be directly used for DMS lane change application due to some reasons.
Firstly, TC is designed for cooperative safety applications \cite{ahmed2011vehicle}, \cite{thorn2018framework}, which are different from the lane change application in consideration here.
Secondly, in lane change application within DMS, it is important to detect both straight and curved roads for TV recognition. However, it is still challenging for existing TC approaches in literature to achieve good performance for curved road geometry.

\begin{figure}[t]
\centerline{\includegraphics[clip,width=.495\textwidth]{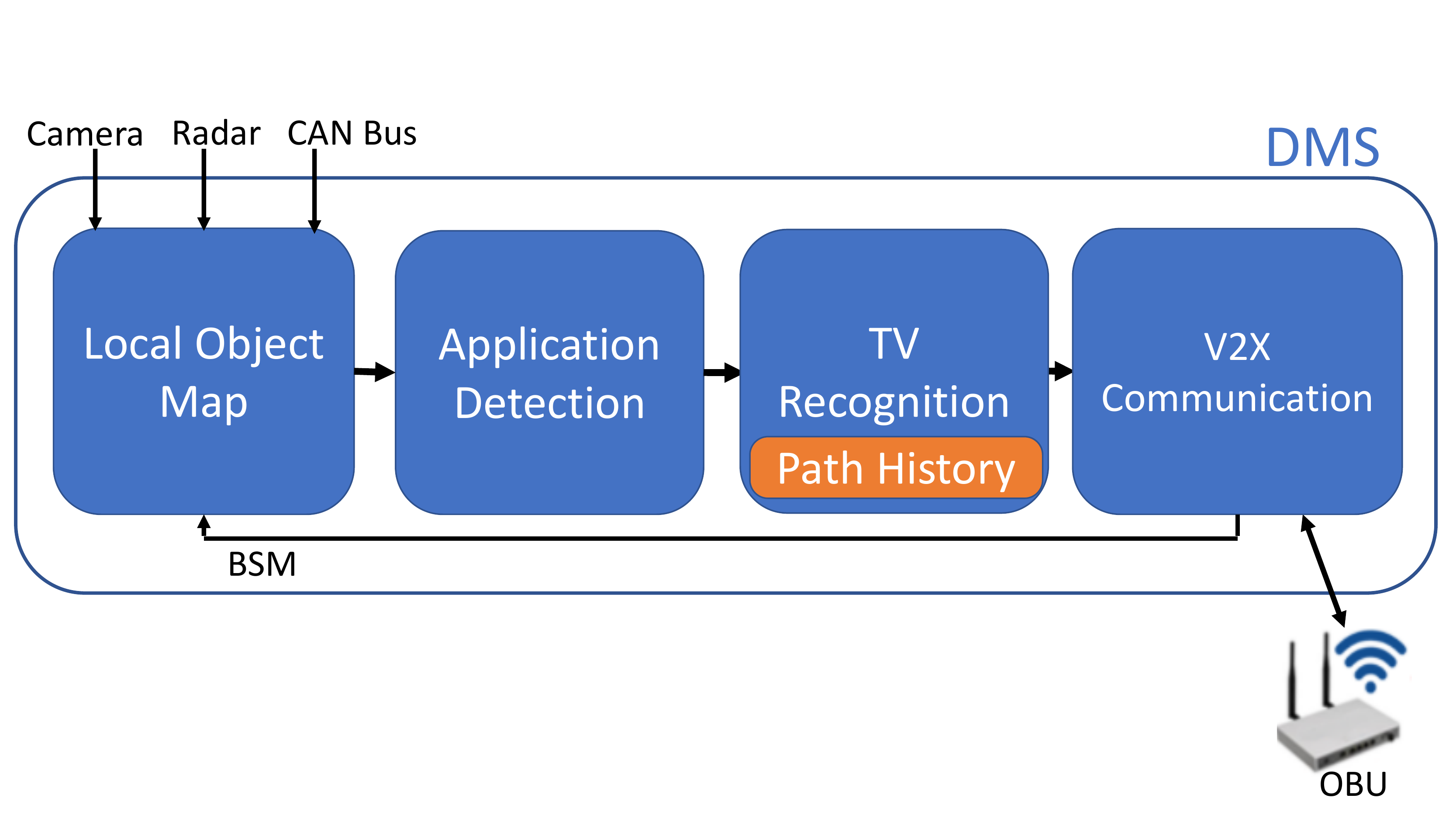}}
\caption{DMS Framework}
\label{fig1}
\end{figure}

\section{DMS Overview}
Figure \ref{fig1} shows a high-level framework of DMS from the perspective of the HV.
The proposed DMS is designed to communicate with the potential TVs regarding the HV's upcoming application-specific intentions and therefore, reduce the number of human error-based crashes.
DMS is designed to utilize the continuously updating local object map containing basic safety information of the nearby vehicles and use it to automatically identify one or more of the pre-defined common 
traffic applications.
Once the scenario is detected, DMS then identifies potential TV(s) and sends a DIM to it informing the TV about the HV's intention.

As shown in Figure \ref{fig1}, DMS is essentially divided into four main components: local object map, application detection, TV recognition, and V2X communication, as detailed below.  

\subsection{Local Object Map}
Local object map module is the initial component of the DMS within a HV. It takes all the sensor, local CAN Bus, and V2X communication information as input and forms a map containing 
information of the nearby vehicles and the road layout. 
The sensor information comprises primarily of front, back, and corner radars and cameras. 
Aside from sensor data, DMS uses V2X communication as well to acquire basic safety information from nearby connected vehicles. Once DMS within the HV receives all the required information, it creates a map by fusing all the received information. The resulting information 
is
where HV can gain access to collective information of its surroundings 
for all its future actions. This map would be subject to updating as soon as new sensor and/or BSM information from any nearby vehicle is available.

\begin{figure}[t]
\centerline{\includegraphics[clip,width=.90\textwidth]{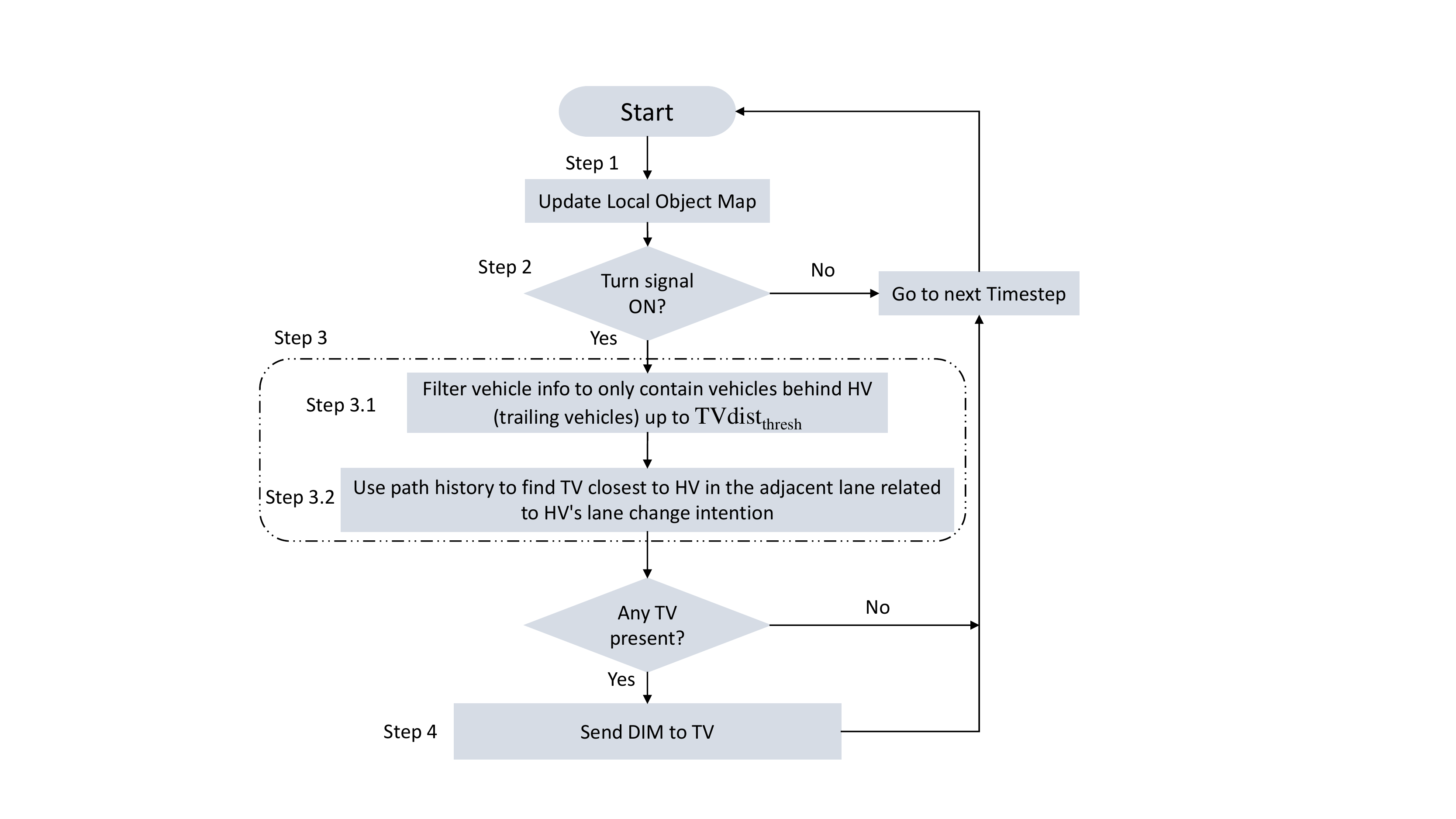}}
\caption{Flowchart for Lane Change Application in DMS}
\label{fig2}
\end{figure}

Since the main focus of this paper is on TV recognition algorithm for lane change application, which is explained later, we assume that all participating vehicles are connected and that the local object map is solely created and updated by BSM information received from the nearby connected vehicles. In other words, we are not complementing the local object map with sensor data for this paper, although it is a part of upcoming studies related to this topic.

\subsection{Application Detection}
Application detection is an essential component of DMS since it is responsible for using the local object map 
to determine whether the HV is in one of the pre-defined traffic applications at any given time. 
Some of the common applications 
considered
within DMS for this project 
are as following:
\begin{itemize}
    \item Lane Change
    \item Ambiguous Right-of-Way at Stop-Sign Intersections
    \item Slow Traffic Ahead
    \item Tailgating
    \item Late Start at a Green Light
\end{itemize}

The focus of this paper is on the lane change application.
Figure \ref{figlc_scenario} represents the DMS process in a typical lane change.
A lane change within DMS can be described as the situation where the HV intends to change its lane from its current lane to an adjacent target lane. 
This situation is triggered as soon as the application detection block receives and checks the updated local object map to detect that the turn signal information from HV's CAN Bus indicates a HV intention to change its lane towards left/right.

\begin{figure}[t]
\centerline{\includegraphics[trim={7.0cm 0 8cm 0},clip,width=0.495\textwidth]{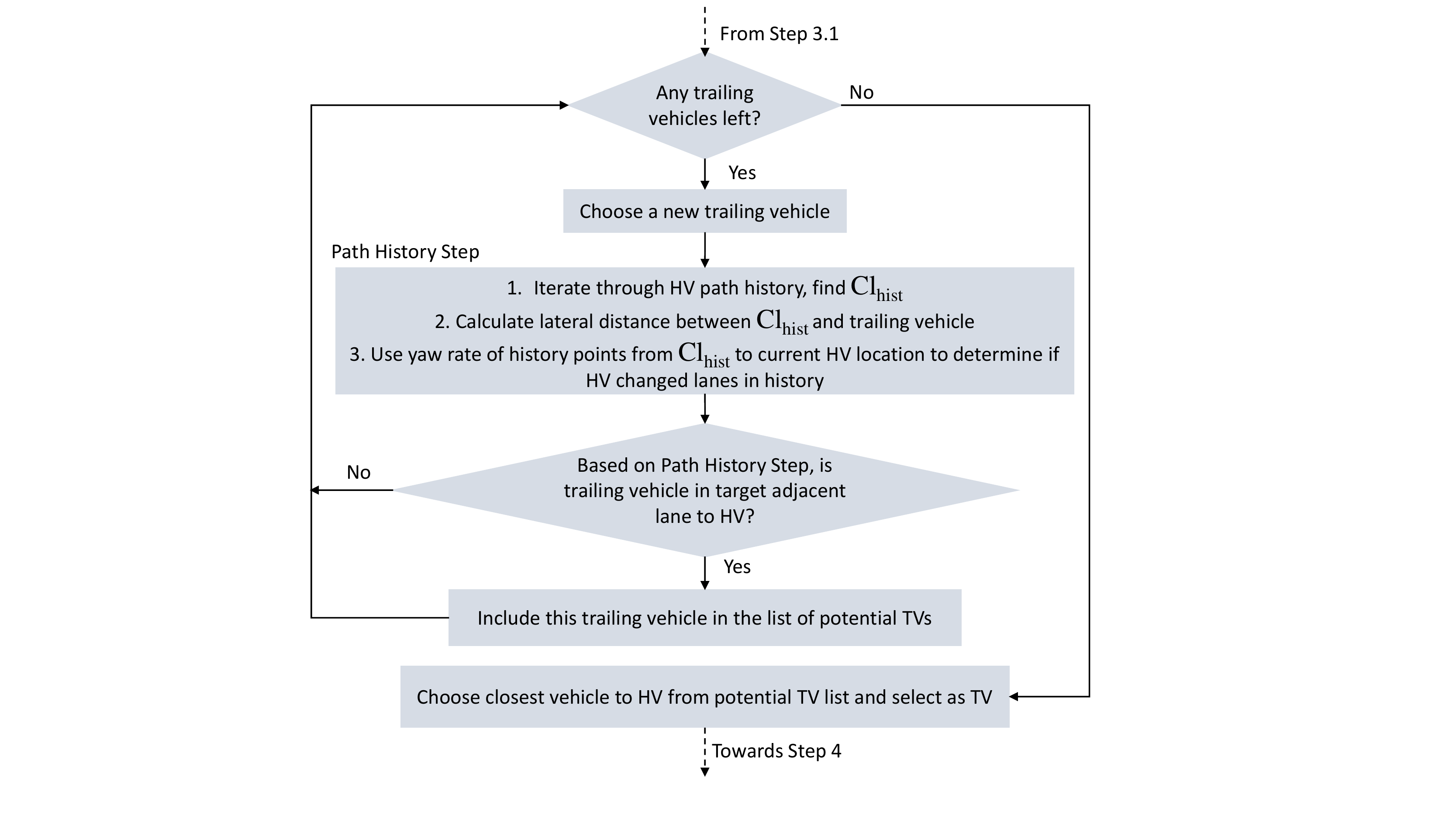}}
\caption{Flowchart for TV Recognition in Lane Change}
\label{fig3}
\end{figure}

\subsection{TV Recognition}
Once the DMS goes past the application detection block and identifies an application (lane change in this regard) from the set of configured DMS applications, the next main objective is to identify the potential TV(s) for the application in consideration. This step is the most significant part of the entire DMS process since the ultimate goal of this system is to serve as an advanced safety feature and inform nearby vehicles of the HV's intention in particular applications. Thus, detection of the correct TV(s) is essential for the desired operation of DMS.

Figure \ref{fig2} provides a high-level understanding of the steps involving lane change within DMS, and then Figure \ref{fig3} specifically highlights the proposed TV recognition algorithm for the lane change application.
It can be seen in Figure \ref{fig2} flowchart that a typical lane change process within DMS is divided into four main steps, as corresponding to the DMS framework described earlier. Step 1 simply involves receiving the updated local object map. Step 2 includes the application detection block where DMS detects HV's lane change intent from the map, and if so, it leads to the triggering of TV recognition block which constitutes Step 3. Finally, if a relevant TV is found, the V2X communication block is invoked as Step 4 where DMS within HV sends a DIM to the TV.

\begin{figure}[t]
\centerline{\includegraphics[clip,width=.495\textwidth]{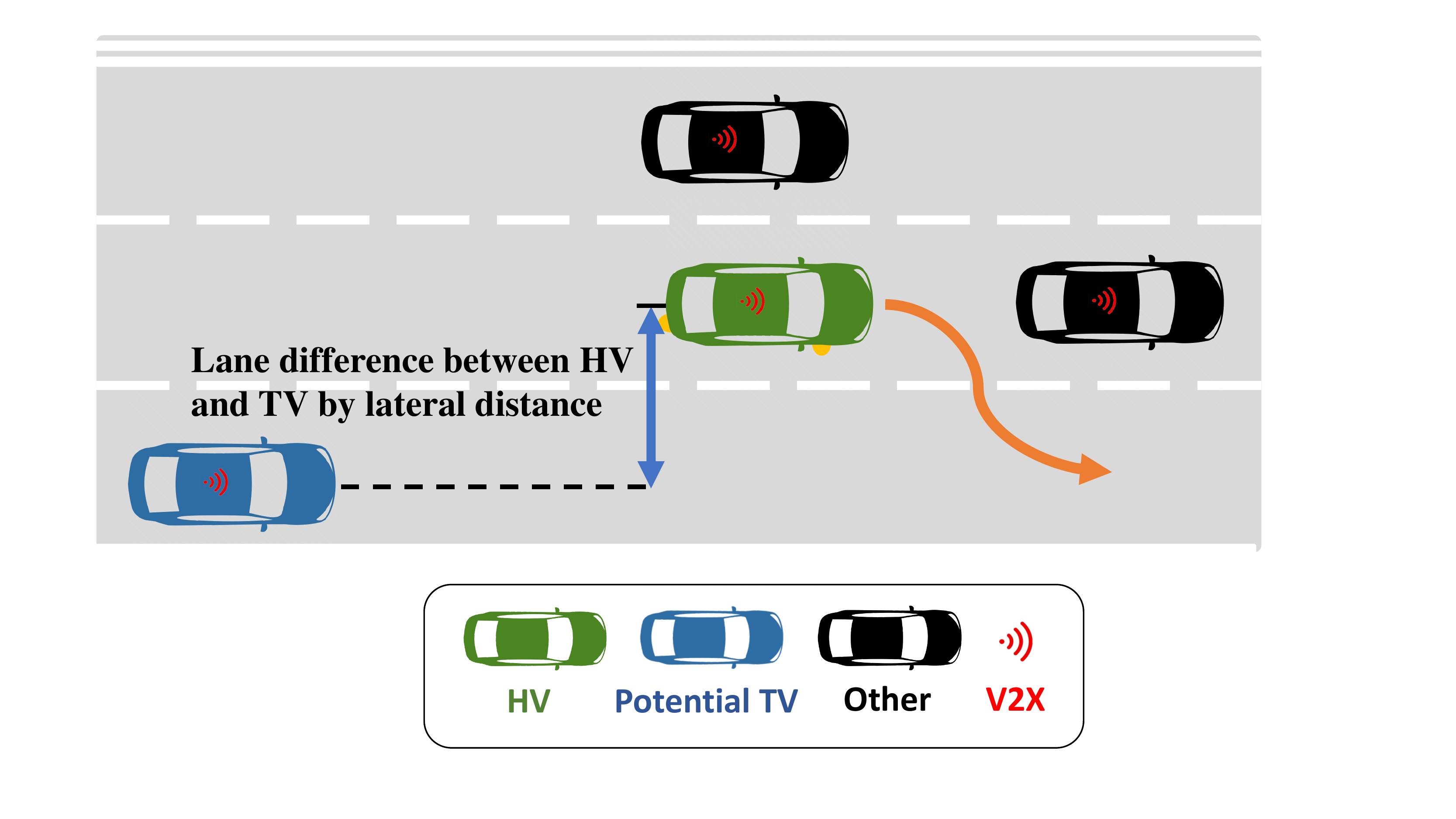}}
\caption{TV Recognition in DMS Lane Change using only Lateral Distance and no PH}
\label{fig_lcscenario_latdist}
\end{figure}

The TV recognition block shown as step 3 in Figure \ref{fig2} comprises mainly of two sub-steps: Step 3.1, and Step 3.2. In Step 3.1, 
DMS analyzes the entire local object map and based on HV's current heading, identifies the trailing vehicles up to a TV distance threshold from the HV (TVdist\textsubscript{thresh}) using longitudinal distances. TVdist\textsubscript{thresh} is a configurable parameter for lane change within DMS and it determines how far DMS can be used to send DIMs to TVs.
Once the vehicles within TVdist\textsubscript{thresh} have been classified as trailing vehicles in Step 3.1, the main objective is to then determine if one of the trailing vehicles can be termed as the TV, as done in Step 3.2. 

The flowchart in Figure \ref{fig3} refers to the details of Step 3.2. This step considers all trailing vehicles to HV and uses the notion of HV's PH to determine if any of the trailing vehicles can qualify as a potential TV.
It should be noted that since lane change application within DMS only considers trailing vehicles to HV, thus HV's PP is not utilized during the TV recognition algorithm.
TV can be formally termed as the trailing vehicle closest to the HV in the adjacent lane that is the target of the HV's lane change intention. This is the vehicle that would directly benefit 
as a result of receiving a DIM from the HV where the HV can inform the TV of its lane change intent.
The notion of HV's PH and the logic behind its usage in TV recognition is further explained in the subsection below.

\begin{figure}[t]
\centerline{\includegraphics[clip,width=.495\textwidth]{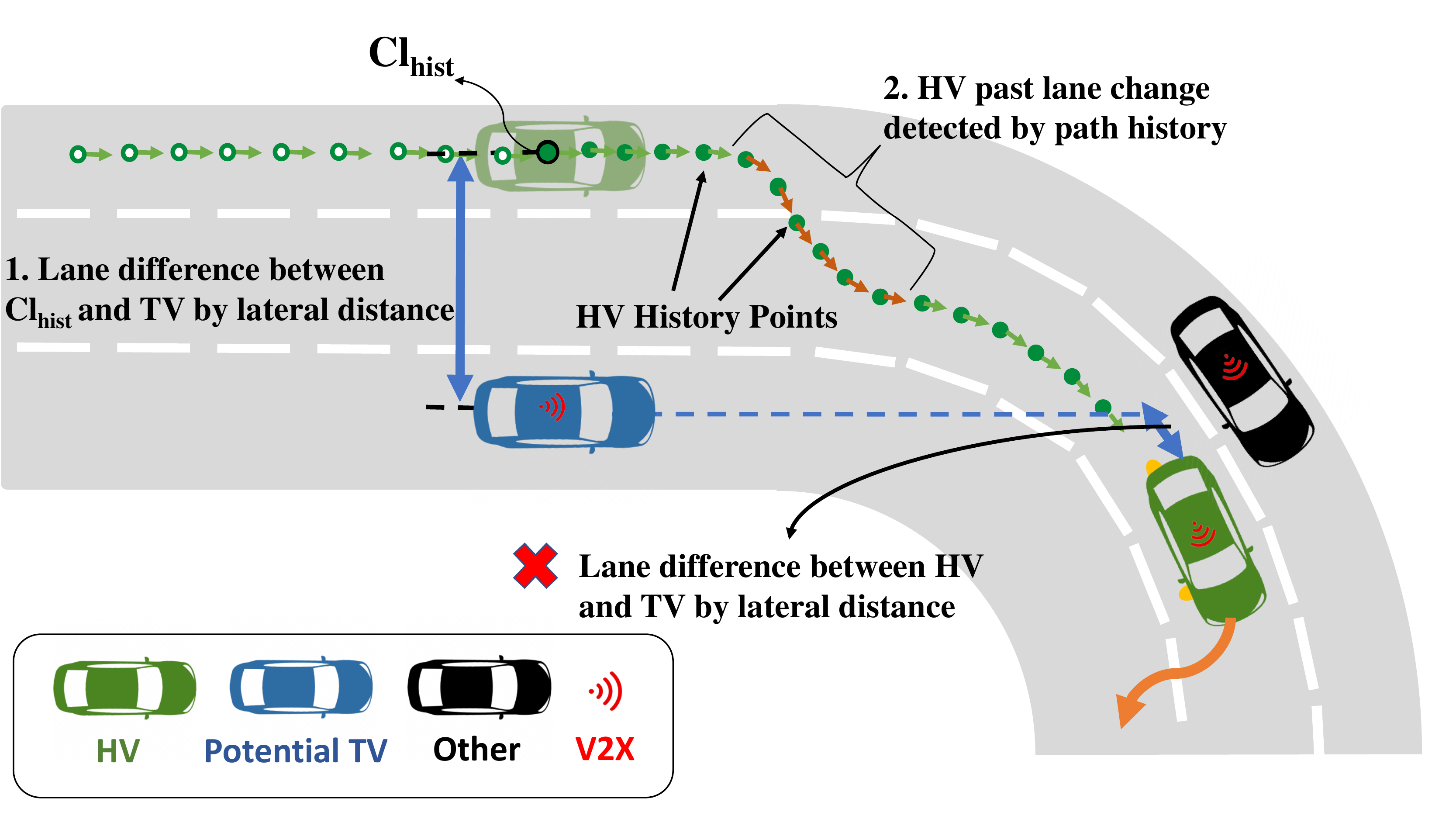}}
\caption{TV Recognition in DMS Lane Change using PH}
\label{fig_lcscenario_ph}
\end{figure}

\subsubsection*{HV Path History}
Choosing a TV solely based on the lateral distance between HV's and a trailing vehicle's current locations can be a viable TV recognition method for lane change application in the cases where the road topology can be characterized as a straight road as shown in Figure \ref{fig_lcscenario_latdist}. However, in the cases of curved roads, lateral distance can yield false readings. To counter this, this study utilizes the notion of HV's PH \cite{sae2016board} as shown in Figure \ref{fig_lcscenario_ph}, that can be used to accurately identify which trailing vehicles are adjacent to HV and thus, be qualified as a potential TV. 
In the case of DMS lane change application when the HV needs to recognize one of the trailing vehicles as a TV in Step 3.2, DMS iterates through HV's PH and chooses the closest HV history point (Cl\textsubscript{hist}) to TV's location. After finding (Cl\textsubscript{hist}), the next step is to calculate the initial lateral distance between (Cl\textsubscript{hist}) and trailing vehicle's current location. 
The initial lateral distance 
determines the lane difference between the trailing vehicle in consideration and HV when the HV was exactly at (Cl\textsubscript{hist}) in the most-recent past. 

In addition to initial lateral distance between (Cl\textsubscript{hist}) and the trailing vehicle, it is also significant to consider the history points between (Cl\textsubscript{hist}) and HV's current location to determine if HV ever performed a lane change in the past between those history points, since this can impact the final lane difference between HV's and trailing vehicle's current locations. Therefore, after finding lateral distance between (Cl\textsubscript{hist}) and trailing vehicle, 
DMS iterates through the history points from (Cl\textsubscript{hist}) to the HV's current location and uses HV's past yaw rates and speeds to determine if there was any lane change by the HV that should be taken into account. 
Thus, as a result of using HV's PH, DMS can detect TVs in the cases of straight and curved roads. 



\subsection{V2X Communication}
The final module in the DMS procedure is the V2X communication block. 
The V2X module in DMS is responsible for multiple important tasks. First, it transmits/receives BSM information to/from nearby vehicles for situational awareness. This information is utilized in the updating of the local object map.
The V2X module also sends and receives DIMs for corresponding DMS applications. In the case when the HV is in a lane change scenario, the TV recognition block instructs the V2X module to send DIM to the identified TV. Similarly, for cases where a vehicle is designated as TV, DIM(s) received from HV is decoded in this same block to allow for application-specific message content.

\section{Analysis and Results}

In this section, we first attempt to test the TV recognition algorithm for lane change application within DMS through a real-world dataset, and then in a simulation setting.

\begin{table}[t]
\centering
\caption{SUMO SIMULATION PARAMETERS}
\begin{tabular}{l r}
\hline
\hline
Experiment Duration             & 20000\text{s}\\
No. of Vehicles                 & 23\\
Vehicle Speed                   & 22, 36 \text{m/s}\\
Vehicle Acceleration
    & 3.5, 7 \text{m/s\textsuperscript{2}}\\
TVdist\textsubscript{thres}         & 50, 75, 100, 150 \text{m}\\
HV Lane Change Intent Periodicity    & 30s\\
Car-Following Model                & Krauss\\
Road Topology                   & Octagonal\\
Number of Lanes
    & 3\\
\hline
\end{tabular}
\label{table:configs_sumo}
\end{table}
%

%
\subsubsection{Real-World Dataset for Benchmarking}
%


In order to test the performance of DMS TV recognition algorithm in lane change application for benchmarking validation purposes,
a dataset is obtained from an experiment conducted on the Interstate-275 near Novi, Michigan. 
The dataset includes two test vehicles that are equipped with On-Board Units (OBUs) which possess GNSS module with acceptable performance.
To test the strength of TV recognition algorithm, the dataset is trimmed to only include data corresponding to a portion of the highway that comprises of a continuous curve 
and contains two lanes where the two afore-mentioned vehicles are driven for about 4 minutes. Throughout the data collection, HV remains ahead of the TV and on the adjacent right lane to that of the TV, and both vehicles do not change their respective lanes. Both the vehicles have a speed of around 30m/s and 
remain within a distance of 200-300m throughout the dataset.

To facilitate the triggering of lane change application within DMS in order to test the proposed logic, a manual left-turn signal is turned on every 10 seconds and the TV recognition results from DMS are stored. Every time the lane change application is triggered, the TV recognition results from the experiment fall into one of the following four categories:
\begin{itemize}
    \item True Positive (TP): Case when DMS correctly detects a TV in the adjacent target lane of HV (TV present in reality and detected).
    \item False Positive (FP): Case when DMS falsely detects a TV in the adjacent target lane of HV (TV not present in reality but detected).
    \item True Negative (TN): Case when DMS correctly does not detect a TV in the adjacent target lane of the HV (TV not present in reality and not detected).
    \item False Negative (FN): Case when DMS falsely does not detect a TV in the adjacent target lane of HV (TV present in reality but not detected).
    
\end{itemize}

The TV recognition algorithm works perfectly throughout the dataset with 100\% TP rating since for every lane change application trigger via turn signal, DMS within HV correctly identifies the trailing vehicle as TV using HV PH as introduced earlier. 
The obtained results from the dataset serve as a benchmark validation for the TV recognition algorithm since the algorithm works correctly under curved road scenario in the dataset and perfectly identifies TVs.


\subsubsection{Simulation Study in SUMO}

Since the TV recognition algorithm performs correctly under a real-world dataset as shown in the previous subsection, the next step is to test 
it
in a simulation setting. For this purpose, we use the Simulation of Urban Mobility (SUMO)
\cite{SUMO2018}
for network building and algorithm testing. SUMO is a microscopic and continuous multi-modal traffic simulator that is capable of handling large traffic networks.
The parameters for SUMO simulation are shown in Table \ref{table:configs_sumo}. These parameters can be easily configured for a simulation in any different setting. 

Any SUMO simulation for this study consists of 1 HV and 22 remote vehicles. 
To create a variety of different TV recognition scenarios during HV lane change intentions,
the simulated vehicles are either configured to possess a maximum speed of 22 m/s or 36 m/s. The accelerations corresponding to these vehicles are 3.5 m/s\textsuperscript{2} or 7.0 m/s\textsuperscript{2}, respectively.
In default SUMO settings, the car-following model utilized is the Krauss model
\cite{krauss1997metastable}
, which is based on the `safe speed' that the trailing vehicle should maintain for safe driving. 
The road topology used for sumo in this study comprises of an octagonal shape consisting of circular and straight road segments.
Each road segment is about 80m long. 
The shape ensures that vehicles can drive in a loop around the topology throughout the simulation duration and can be tested for multiple different straight and/or curved road situations.


As explained in previous sections, one of the main contributions of DMS in the lane change application is that it can allow the TV to receive advanced information regarding the HV's upcoming lane change in front of it, and thus, allow the TV enough time to react to it and reduce any chance of a near-collision situation with the HV. In the case of SUMO simulation for this study, this reaction by TV is translated into a braking maneuver performed by TV 
once it receives a DIM from the HV regarding the latter's intention of lane change.
The braking maneuver in SUMO is represented by reducing the TV's current speed by a moderate 3 meters/second, although it can be altered for future studies.
At the same time as the TV presses brakes in response to DIM, the HV performs a lane change maneuver and gets into the same lane as TV. This allows an increase in the inter-vehicular distance between the HV and the TV which translates to an increase in time and space heading, as discussed later. In a similar lane change situation in SUMO without DMS, the HV performs a lane change but since TV does not receive a DIM from HV, it continues to follow the default SUMO car-following model and may or may not use brakes depending on its distance and speed difference to the HV. 
Therefore, HV and TV can come closer to each other in lane change scenarios without DMS.
Both the braking maneuver by TV and lane change maneuver by HV requires an on-line interaction with SUMO whenever a lane change situation occurs. To do so, the authors utilize Traffic Control Interface (TraCI) \cite{wegener2008traci}.


\begin{figure*}[t]
    \centerline{\includegraphics[width=0.9\textwidth,height=9cm]{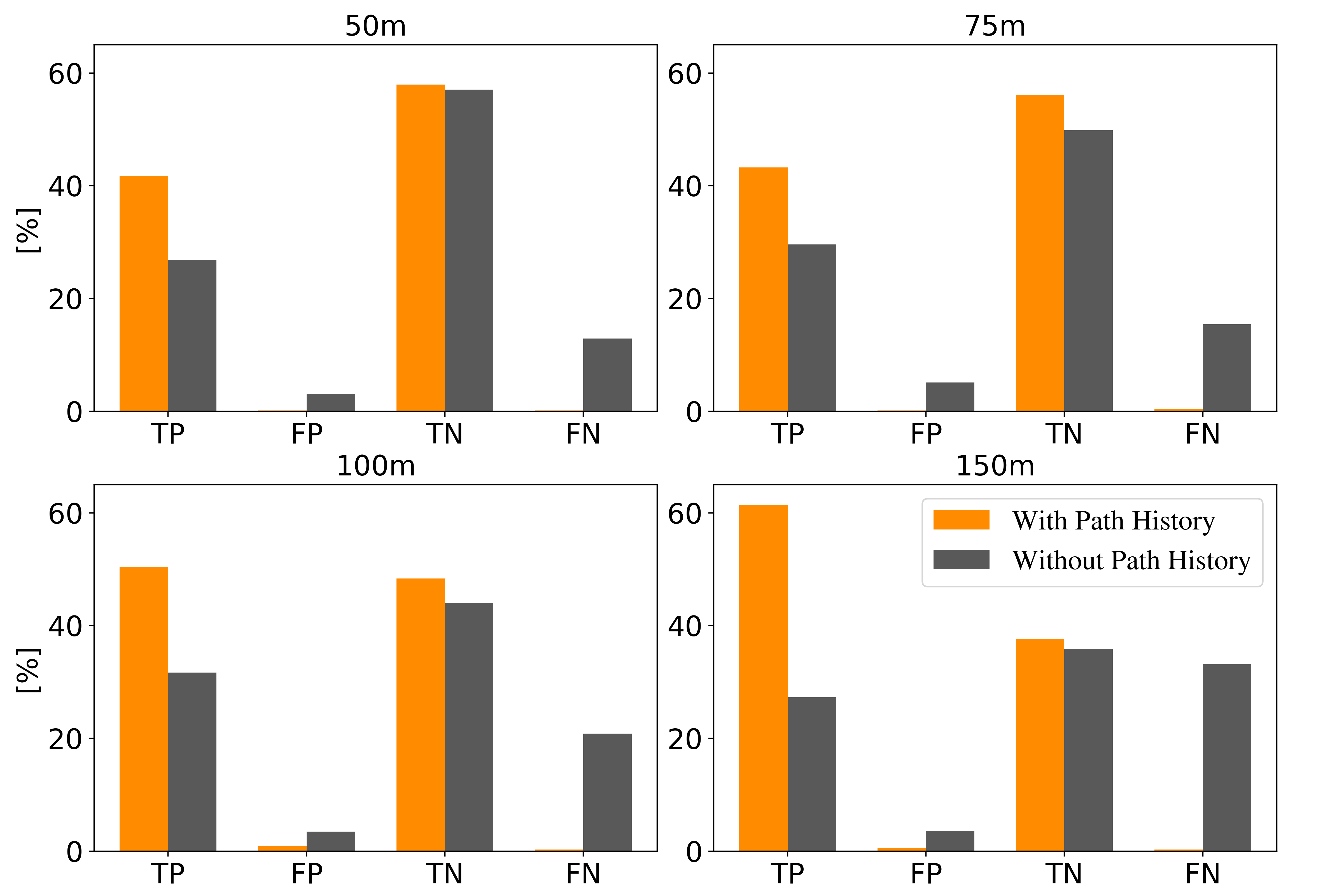}}
    \caption{TV Recognition Statistics in SUMO using DMS with and without PH}
    \label{lc_tvrec_full}
\end{figure*}

In order to test the TV recognition algorithm and show the impact of DMS, a total of four different TVdist\textsubscript{thresh} have been used, i.e. 50, 75, 100, and 150 m. The higher the TVdist\textsubscript{thresh}, the larger can be the distance between HV and TV, and thus the greater the reliance of DMS on the PH for TV recognition. For each TVdist\textsubscript{thresh}, three different simulations have been run: 
\begin{itemize}
    \item First simulation considers SUMO scenario with DMS using PH for TV recognition.
    \item Second simulation considers DMS with only lateral distance calculation between HV and trailing vehicles for TV recognition and no PH.
    \item Third simulation assumes no DMS and just default SUMO behavior pertaining to the car-following model.
\end{itemize}
  Thus, a total of 12 SUMO simulations are run for this study.
  It should be noted that in the third simulation for every TVdist\textsubscript{thresh}, whenever the HV possesses a lane change intention towards a lane containing a trailing vehicle, the HV proceeds with the lane change and since the trailing vehicle does not receive DIM, it continues to proceed towards HV until car-following model gets activated, as explained earlier.

Figure \ref{lc_tvrec_full} shows TV recognition statistics comparison between first (scenarios using DMS with PH) and second (scenarios using DMS with lateral distance and no PH) simulations for all TVdist\textsubscript{thresh}. At a high level, it can be clearly observed that scenarios with PH outperform those without PH at all TVdist\textsubscript{thresh}. Looking specifically at TP and TN statistics, scenarios with PH have higher TP and TN than scenarios without PH. As TVdist\textsubscript{thresh} rises, TP of scenarios with PH can be seen to rise and TN to fall. This shows the impact of PH since an increase in TVdist\textsubscript{thresh} allows DMS to detect more TVs and almost all detections are correct, leading to a higher TN and negligible FP and FN readings. On the other hand, in the scenarios without PH, as TVdist\textsubscript{thresh} increases, the TP can be seen to be more or less stagnant with a minor reduction in TN but a noticeable increase in FN. This highlights the drawback of just using lateral distance for lane difference calculation between HV and trailing vehicles since a higher distance between HV and trailing vehicles can lead to higher source of error in lane difference calculation, specially in the cases of curved roads. FP percentages of scenarios without PH can also be observed to be significantly higher than those with PH.

\begin{figure*}[t]
    \centerline{\includegraphics[width=0.88\textwidth, height=9cm]{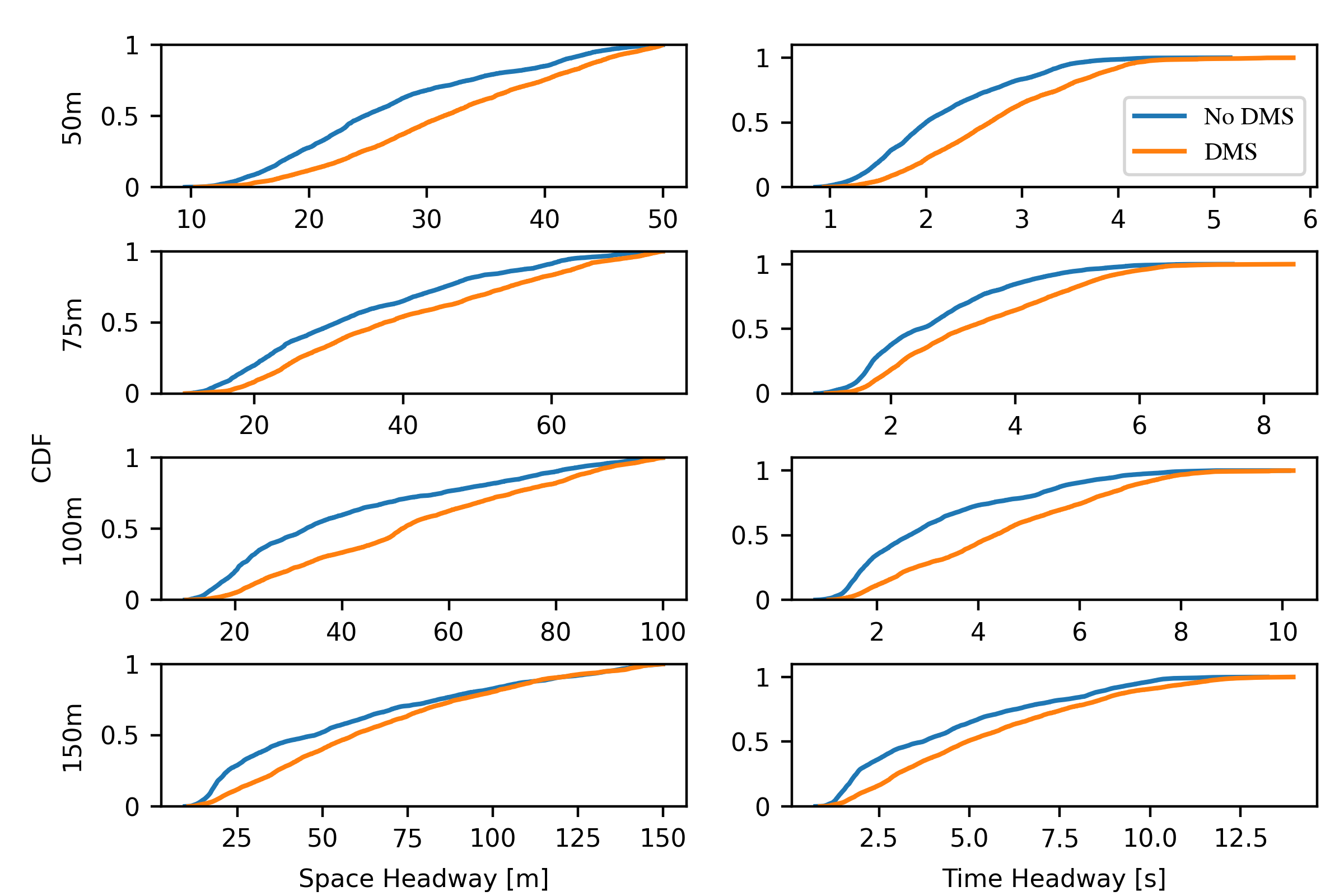}}
    \caption{Time and Space Headway Comparison between DMS with PH and No DMS}
    \label{lc_dms_nodms}
\end{figure*}

In addition to TV recognition algorithm comparison to show the benefits of using PH, SUMO simulations are also run to show the impact of using DMS in terms of advanced safety. Therefore, 
in line with the reaction model for HV and TV in response to a lane change scenario with or without DMS as explained earlier, this study uses two metrics to show the impact of DMS: Time and Space Headway. Time headway is defined as the time difference between HV and TV when they are in the same lane after HV changes its lane. On the other hand, space headway is defined as the longitudinal distance between HV and TV when they are in the same lane after HV changes its lane.
To show an unbiased and holistic impact of DMS, the time and space headway values are recorded for a specified duration, i.e. 10 seconds, immediately after every HV lane change.

Figure \ref{lc_dms_nodms} shows time and space headway comparison between first (scenarios using DMS with PH) and third (scenarios without DMS) simulations for all TVdist\textsubscript{thresh}. At a high level, it can be observed that as a result of using DMS, both space and time headway metrics outperform scenarios not using DMS. This can be simply understood by the reaction model where DMS allows TVs to receive DIMs from HV, thereby allowing them to brake and hence, increase the heading numbers, as opposed to scenarios without DMS, which solely rely on the car-following model. The space heading difference for all TVdist\textsubscript{thresh} can be seen to be consistently higher for DMS scenarios specially at distances within 100m since that is where braking can result in a bigger difference in headway values and reduce near-collision situations. Similarly, time headway values in DMS scenarios also show an increase in time due to braking maneuver which can prove to be very beneficial specially in low-distance situations between HV and TV since the distance and relative speed between HV and TV may not be small enough to force the operation of car-following model but it can nevertheless be sufficient enough to take an advanced safety action of braking and increase the distance between TV and incoming HV.

\section{Concluding Remarks}
This study introduces DMS for cooperative safety scenario-specific applications with an emphasis on lane change application. 
DMS uses BSMs from nearby vehicles to update the HV's local object map. This map is then used to identify potential TVs in the adjacent lane corresponding to the HV's turn signal direction. Once the TV has been identified, DMS within HV sends an OTA DIM to the TV regarding HV's upcoming lane change intention.
Main focus is on the proposed TV recognition algorithm with PH for lane change application. The paper explains the operation of the algorithm in both straight and curved road trajectories.
The TV recognition algorithm is tested using a real-world dataset and also SUMO traffic simulator 
and its performance in both platforms confirms the accurate functionality of the algorithm.

\section{Future Works}


Firstly, it will be beneficial to implement DMS in a real vehicle to compare the proposed TV recognition algorithm with a similarly designed TC-based lane change application for a realistic baseline.
Secondly, the TV recognition algorithm for lane change can be tested using more sophisticated and complex road scenarios in simulation platforms such as CARLA.
Thirdly, the algorithm can be further tested in scenarios including non-connected vehicles and communication losses. For this step, we can aim to leverage sensor fusion using camera and radar in addition to BSMs. Finally, as mentioned earlier, we can include more safety applications within DMS framework.

\section*{Acknowledgment}
We are grateful to Hossein Mahjoub for his help on providing the dataset used for benchmarking, and thanks to Hidekazu Araki for the discussions on road curvature edge cases.
The contents of this paper reflect the views of the authors, who are responsible for the facts and the accuracy of the data presented herein. The contents do not necessarily reflect the official views of Honda.

\bibliography{main.bib}{}
\bibliographystyle{unsrt}

\end{document}